\documentstyle[aps,psfig,manuscript]{revtex}
\begin{document}
\draft

\title{ Strain Relaxation Mechanisms and Local Structural Changes
in Si${}_{1-x}$Ge${}_{x}$ Alloys}

\author{Ming Yu, C. S. Jayanthi, David A. Drabold${}^{\ddagger}$, and S. Y. Wu}
\address{Department of Physics, University of Louisville,
Louisville, KY, 40292\\
${}^{\ddagger}$ Department of Physics and Astronomy, and
Condensed Matter and Surface Sciences Program,
Ohio University, Athens, OH 45701-2979}


\maketitle

\begin{abstract}

In this work, we address issues pertinent to the
understanding of the structural and electronic properties of
Si${}_{1-x}$Ge${}_{x}$ alloys, 
namely, (i) how does the lattice constant mismatch between bulk 
Si and bulk Ge manifests itself in the alloy system? and (ii) what are 
the relevant strain release mechanisms? To provide
answers to these questions, we have carried out an in-depth study of the
changes in the local geometric and electronic structures arising
from the strain relaxation in Si${}_{1-x}$Ge${}_{x}$ alloys.  We first
compute the optimized lattice constant for different compositions
($x$) by fully relaxing the system and by minimizing the total
energy with respect to the lattice constant at each composition,
using an {\it ab initio} molecular dynamics scheme.
The optimized lattice constant, while exhibiting a general trend
of linear dependence on the composition (Vegard's law), shows a
negative deviation from Vegard's law in the vicinity of $x$=0.5. 
We delineate the mechanisms responsible for each one of the above
features. We show that the radial-strain relaxation through bond
stretching is responsible for the overall trend of linear
dependence of the lattice constant on the composition. On the
other hand, the negative deviation from Vegard's law is shown
to arise from the angular-strain relaxation. More specifically, the
combined effect of the local bond-angle deviations from the tetrahedral 
angle and the magnitudes of the corresponding peaks for the partial-angle 
distribution function determines the negative deviation from Vegard's law.
The electronic origin of the changes in the local geometric structure 
due to strain relaxation is also
presented in this work. In particular, the correlation between the
bond charges and the bond-lengths for Si-Si,
Ge-Ge, and Si-Ge pairs in Si${}_{1-x}$Ge${}_{x}$ alloys for
different compositions is explicitly shown. Our calculation of the 
average coordination number as a function of composition indicates 
a random occupation of Si and Ge on the lattice sites, suggesting that 
Si and Ge atoms are fully miscible in the alloy system. 

\end{abstract}


\newpage
\section{Introduction}

It is well-known that the 4\% difference in the experimental observed
lattice constants between bulk Si and bulk Ge gives rise to a significant strain
in the growth of Si${}_{1-x}$Ge${}_{x}$ alloys, and the relaxation of the strain
causes changes both in the local geometric structure and in the electronic 
structure of Si${}_{1-x}$Ge${}_{x}$ alloys as compared to those of pure Si and Ge. 
In order to understand the mechanism of the strain relaxation, extensive experimental
\cite{incocia,nishino,kajiyama,matsuura,woicik,aldrich,wei,hitchcock,aubry,expt,dismukes} 
and theoretical
\cite{martins,weidmann,ichimura,gironcoli,qteish,cai,mousseau,landau} 
efforts have been devoted to the study of the interplay between the relaxation of
the strain and the changes in local geometric and electronic structures. 
Most experiments \cite{kajiyama,aubry,expt,dismukes} found that the lattice constant as a function of the composition
does not follow an exact linear relation such as the one given by
the Vegard's model \cite{vegard} but has a negative deviation from the
Vegard's law. The bond lengths, on the other hand, show a
weak composition dependence
\cite{incocia,nishino,matsuura,woicik,aldrich,wei,hitchcock}. But they do not obey the Pauling
model \cite{pauling} in which the bond length between a pair 
of atoms is independent of composition, and the steric strain in the 
alloys is accommodated by bond-angle changes. 

A number of theoretical studies have been devoted to the local
structural analysis of Si${}_{1-x}$Ge${}_{x}$ alloys at an empirical 
or at the semi-empirical level. Weidmann and Newman
\cite{weidmann} by minimizing a model strain energy function 
found that the bond lengths between Si-Si, Si-Ge,
and Ge-Ge pairs as a function of composition are straight lines
and parallel to each other. Similar result were also obtained by Ichimura 
{\it et al.} \cite{ichimura} and Gironcoli {\it et al.}
\cite{gironcoli}.
Alternatively, Thorpe and co-workers \cite{cai,mousseau}
proposed a simplified
model on the basis of macroscopic elastic properties.
The composition dependence of the bond lengths is described
via a topological rigidity parameter $a^{\ast \ast}$ which
leads to the Vegard limit when  $a^{\ast \ast}$=0 and to the
Pauling limit when  $a^{\ast \ast}$=1. According to their model,
$a^{\ast \ast}$ should be 0.707 
for SiGe alloys, and a plot of
the Si-Si, Si-Ge, and Ge-Ge bond lengths versus composition
should consist of three equally spaced parallel lines
having a slope that is directly related to the value of 
$a^{\ast \ast}$.

These previous theoretical studies, while attempting
to provide the insight into how the local structural
properties accommodate the relaxation of the strain,
failed to predict the weak dependence of the bond
lengths of Si-Si, Si-Ge, and Ge-Ge on the composition.
They also did not reproduce the negative deviation
of the lattice constant from the Vegard's law.
For example, the result given in Ref. \cite{mousseau}
predicted a linear dependence on the composition
for the lattice constant while the result by a Monte
Carlo simulation \cite{gironcoli} yielded a positive deviation from
Vegard's law. Another Monte Carlo study using a
statistical-mechanical model \cite{landau} also obtained an
overall linear dependence of the lattice parameter on the
composition, but with a hint of negative deviation from Vegard's
law in the vicinity of $x \simeq 0.5$.
An effort to resolve these issues
had been carried out by Shen {\it et al.} \cite{shen},
using a semi-empirical tight-binding method in the
dilute limit. The model described reasonably  well the behavior
of the lattice constant and the properties of bond 
lengths in this dilute limit. However, no attempt was
made to correlate the model of strain relaxation
and the local properties in this study.

Very recently, a more accurate experimental measurement
of the local structure at all compositions has been reported
by Aubry {\it et al.} \cite{aubry}. They analyzed the K-edge X-ray
absorption fine structure (XAFS) spectra of Si and Ge in strained and
relaxed Si${}_{1-x}$Ge${}_{x}$ alloys. They found
that the Si-Si, Si-Ge, and Ge-Ge first-shell distances
show a weak dependence on the composition. The slopes of the linear fits 
to the bond lengths as a function of the composition are demonstrably 
different from each other \cite{aubry}. This result is different from the previous 
theoretical predictions. They also confirmed from the composition
dependence of the coordination numbers that the Si and Ge atoms are
likely to be randomly occupying the sites and are fully miscible at
all compositions. 

An accurate theoretical determination of equilibrium configurations of
Si${}_{1-x}$Ge${}_{x}$ alloys with no parametric input is highly desirable.
This is because such a determination can help to clarify issues related
to the whole spectrum of available experimental observations.
Furthermore, it can shed light on the interplay between the effects of
strain relaxation and local properties of the alloys.
In this work, we have used the {\it ab initio} molecular
dynamics scheme, as developed by Sankey and co-workers
\cite{sankey}, to carry out the energy minimization for the determination of
the equilibrium structures of Si${}_{1-x}$Ge${}_{x}$ alloys at various
compositions. A brief outline of this method is given in
section II. We then conducted a local analysis of the
structural (section III) as well as electronic properties
(section IV) of the relaxed configurations.
Corresponding to each composition, we have computed the optimized
lattice constant, bond lengths, coordination numbers, and
pair-correlation functions (radial and angular). The results of these 
calculations are used to identify the mechanisms for strain release and 
to explain succinctly the origin of Vegard's law as well as the deviation 
from this law for $x$ in the vicinity of 0.5. The conclusions drawn from this
work are given in section V. 
 
\section{Method}

The {\it ab initio} molecular dynamics scheme employed in the present
work is based on the density-functional theory (DFT) in the
local-density approximation (LDA), as developed by
Sankey and co-workers \cite{sankey}, where a local basis set is
used to construct the Kohn-Sham orbitals. These basis functions
are slightly excited pseudo-atomic orbitals (PAO). The Kohn-Sham
orbitals are calculated self-consistently using the 
Hamann-Schl\"uter-Chiang pseudo-potentials \cite{hamann} and 
the Ceperley-Alder form of the exchange-correlation potential, 
as parameterized by Perdew and Zunger \cite{exc}. The use of PAOs
as basis set is extremely convenient in studies which require
extracting information about local structural and electronic
properties of complex systems such as Si${}_{1-x}$Ge${}_{x}$ alloys. For
complex systems with reduced symmetry, the computational
performance of this method as far as the CPU time is concerned is
better than other {\it ab initio} methods based on plane-wave basis
sets.
 
In our simulation, $sp^{3}$-type PAOs were used with confinement
radii of 5.0 $a_{\rm B}$ and 5.2 $a_{\rm B}$ for Si and Ge atoms,
respectively. The initial network chosen has a tetrahedral
symmetry with 216 atoms in a cubic unit cell. For a given volume
and composition, the network was fully relaxed by the dynamical
quenching method using the quantum-molecular
dynamics scheme cited above. The equilibrium configuration was
considered to have been reached when 
the force on each atom is less than $1 \times 10^{-2}$
eV/\AA. We evaluated the total energy convergence by using both 1
and 8 special $k$ points in the BZ and found that the result of using 8 $k$
points only improves the accuracy by $1 \times 10^{-2} $eV. Therefore, we
adopted the $\Gamma$ point calculation in the ensuing simulations. The volume
optimization was carried out by minimizing
the total energy with respect to the lattice constant for
a given composition. We then conducted a local analysis of the
structural and electronic properties corresponding to each
composition.

\section{Local Analysis of Structural Properties}

Two types of configuration models were considered in our
simulation. The first type is referred as a {\it regular} type.
In this model, the sites in the supercell are first labeled sequentially in a certain
order from site 1 to site $N$ where $N$ is the total number of atoms in the
supercell. The assignments of Si and Ge atoms at the sites are carried out
in a regular pattern according to their concentrations from site 1 through site $N$.
For example, in the case of Si${}_{.75}$Ge${}_{.25}$, the assignment of Si and
Ge atoms at a given site proceeds following the rule of one Ge atom
after every three Si atoms. The second type is referred as {\it random} in 
which the Si and Ge atoms are (completely) randomly distributed. We found that
the total energy difference between these two cases is quite small (within the error
bar), but the coordination numbers as a function of the composition
are quite different as shown in Fig. 1.  It is found that the {\it regular} 
type (open symbols) can not explain the experimental result \cite{aubry} 
(see the inset), but the {\it random} type (solid symbols) mimics the experimental 
result very well. We therefore concentrated on the {\it random} type
hereafter in our local structural analysis and
in the comparison with the experimental results. Other details of
results shown in Fig. 1 will be discussed later after we
introduce the relevant definitions.

We examined the global and local structural properties of 
Si${}_{1-x}$Ge${}_{x}$ alloys at various compositions 
($x=0, 0.10, 0.25, 0.40, 0.5, 0.60, 0.75, 0.90, 
$ and 1.0). Figure 2 presents the optimized lattice constant 
versus the composition obtained from the total energy minimization.
We found that the lattice constant (solid circle) monotonously
increases from 5.513 \AA $\;$ at $x$=0 (corresponding to pure Si) 
to 5.645 \AA $\;$ at $x$=1 (corresponding to pure Ge) with increasing $x$.
It exhibits a negative deviation from
the linear Vegard's law (dashed line). The deviation appears
from $x \approx 0.25$, shows the largest deviation around $x=0.5$, and
then gradually disappears beyond $x=0.75$, in good
agreement with experimental results \cite{aubry,expt} (see the inset).
It should be noted that, while the experimental value of the mismatch between 
bulk Si and bulk Ge is about 4\%, our optimization yields a mismatch of only about 
2.4 \%. Therefore, we present the experimental data of Ref. \cite{aubry}
in the inset rather than in the same figure. Our goal is to
compare the trend and the general pattern of the structural changes as a function
of concentration so as to deduce an understanding of the strain
relaxation mechanisms in Si${}_{1-x}$Ge${}_{x}$ alloys.

The average bond-length between a pair of atoms of types $\alpha$ and
$\beta$, $ b_{\alpha \beta}$, is defined as:
$ b_{\alpha \beta}=\sum_{i_{\alpha}}^{N_{\alpha}}
\sum_{j_{\beta}}^{R_{\rm cut}}
d_{i_{\alpha},j_{\beta}}/\sum_{i_{\alpha}}^{N_{\alpha}}
n_{i_{\alpha}}(\beta)$,
where $\alpha (\beta)$ denotes the type of atom, $N_{\alpha}$ is
the total number of $\alpha$-type atom in the supercell, $d_{i_{\alpha},j_{\beta}}$
the distance between the $\alpha$-type atom at the $i$th site
and the $\beta$-type atom at the $j$th site, and $n_{i_{\alpha}}(\beta)$ 
the number of neighboring $\beta$-type atoms 
around the $\alpha$-type atom at the $i$th site 
within the cutoff radius of $R_{\rm cut}$. We took $R_{\rm cut}$ to be
2.7 \AA $\;$ in our analysis which is between the first and the second peaks of the
radial-pair distribution function of the relaxed alloy configurations. 
We examined  the choice of $R_{\rm cut}$ in the region of 2.6-3.2 \AA $\;$ 
and found that the value does not have much influence on the results because the 
first and the second peaks are well separated by about 1 \AA.

Figure 3 illustrates the calculated average bond-lengths of
Si-Si, Si-Ge, and Ge-Ge pairs versus the composition $x$. 
The experimental data \cite{aubry} are presented in the inset.
The experimental error bars in Ref. \cite{aubry} have not been
shown in the inset because we are only concerned with comparing
the trend exhibited by the theoretical result with that from
the experimental data. The largest error bar for Si-Si bond length
in Si${}_{1-x}$Ge${}_{x}$ alloys occurs at $x \sim 0.75$ while
that for Ge-Ge bond length at $x \sim 0.25$ \cite{aubry}.
These occurrences had been attributed to the distortions associated
with possible compound formation at these concentration \cite{aubry}.
If the data points at these two concentrations are ignored, the
agreement between the trend exhibited by our theoretical
calculation and that shown by the experimental result is indeed
very good. Turning now to the theoretical result,   
it can be seen that the variation of both sets of bond lengths
(Si-Si and Ge-Ge) with respect to $x$ follows the same general pattern. Overall,
the bond lengths are rather insensitive to the composition. 
For the Si-Si pairs, the increase in their average bond-length is
concentrated in the Ge-rich region while for the Ge-Ge pairs, the decrease in their
average bond-length is mostly in the Si-rich region. The Si-Ge
pairs appear to form at a distance close to the mean value of the
average bond-lengths of Si-Si and Ge-Ge pairs for a given composition.
The weak dependence of the average bond-lengths of Si-Si and
Ge-Ge pairs on the composition indicate that the Si-Si and Ge-Ge pairs
prefer to maintain their respective bond-length even in the
alloying situation. The preference of the Si-Ge pairs
to form at distances close to the
mean value of the average bond-lengths of Si-Si and Ge-Ge pairs is an 
indication that the mismatch between lattice constants of bulk Si and bulk Ge
is accommodated by the formation of the Si-Ge bond. 
It should be noted that the bond lengths vs the composition curves for Si-Si, Si-Ge, 
and Ge-Ge bonds obtained in the present calculation do not 
correspond to equally spaced parallel lines as obtained by previous calculations
\cite{weidmann,ichimura,gironcoli,cai,mousseau}. They are, however, 
consistent with the experimental result (see the inset of Fig. 3)
as discussed above. It should also
be noted that the calculated average bond-length of Si-Si pairs 
for bulk Si is 2.39 \AA, somewhat longer than the experimental
value of 2.35 \AA $\;$ while the calculated average bond-length
of Ge-Ge pairs for bulk Ge is 2.44 \AA, somewhat shorter
than the experimental value of 2.45 \AA \cite{dismukes}. Thus the spread
of the variation of the calculated average bond-lengths
versus the composition is narrower than that of the 
corresponding experimentally observed bond lengths.

The coordination numbers versus the composition are presented in Fig. 1.
Note that the coordination numbers are defined as
$N_{\rm average}=\sum_{i}^{N} n_{i}/N$, and 
$N_{\alpha \beta}=\sum_{i_{\alpha}}^{N_{\alpha}(\beta)}
n_{i_{\alpha}}(\beta)/
N_{\alpha}(\beta)$, where $N$ is the total number of atoms, $n_{i}$ 
the number of neighbors of the $i$th atom, and $N_{\alpha}(\beta)$
are the total number of the $\alpha$-type atom having $\beta$-type atoms as
its neighbors within the cutoff $R_{\rm cut }$,
respectively. It is clear that $N_{\rm average} =4$ in the four-fold bonding
structure such as tetrahedral symmetry systems, and $N_{\alpha \beta}=1$
when there is only one $\alpha$ ($\beta$) atom in the $\beta$ ($\alpha$) atom
system. If $\alpha$ and $\beta$ atoms are randomly
distributed, there must be no significant difference
among $N_{\alpha \alpha}$, $N_{\beta \beta}$, $N_{\beta \alpha}$, 
and $N_{\alpha \beta}$ at $x=0.5$. The coordination numbers of the {\it random} type 
clearly show such a behavior as can be seen in Fig. 1 
and our results are consistent with the recent experimental
measurements \cite{aubry}. Thus, our results support the notion that Si and Ge atoms 
randomly occupy the sites and are fully miscible in Si${}_{1-x}$Ge${}_{x}$ alloys
because of their similar chemical properties.

We next analyze how the local structure changes due to the strain
relaxation. The radial-strain relaxation can be analyzed 
from the radial-pair distribution function 
$g_{\alpha \beta} (r)$ for an $\alpha$-type atom
at the origin and a $\beta$-type atom at a distance $r$ away. 
As shown in Fig. 4, the total radial-pair distribution function $g(r)$ 
(solid curve) in the region of the first-shell distance consists of three sub-peaks: 
the Si-Si peak on the left side, the Si-Ge in the
middle, and the Ge-Ge on the right side (note that these
sub-peaks can be clearly distinguished at $x$=0.4, 0.5, and 0.6). Such peak positions
shift less than 0.5\% from $x=0.0$ to $x=1.0$,
indicating that during the strain relaxation all the pairs 
of Si-Si, Si-Ge, and Ge-Ge prefer to be as close to their equilibrium distances 
as possible. It can also be seen that the 2\% 
shift of the peak position of the average 
first-shell distance is attributed to the change of the
ratio among the Si-Si, Si-Ge, and Ge-Ge pairs in the mixture.
This explains the monotonous increase of the lattice constant 
with increasing Ge composition.   
Furthermore, the average overall bond-length of all three types of bonds
has been plotted versus the composition in Fig. 3. It shows a linear
dependence on the composition. Hence the strain release
through the radial stretching is apparently responsible for the
overall trend of linear dependence  of the lattice constant on $x$.

The question is then how does the lattice constant versus $x$ curve exhibit a 
negative deviation from Vegard's law? Specifically, how does the mismatch
between the lattice constants of bulk Si and bulk Ge manifest itself
when Si-Si and Ge-Ge pairs prefer to maintain 
their respective lengths? To answer this question, we examined 
the bond-angle strain relaxation from the bond-angle
distribution function $g(\theta)$. It is seen from Fig. 5 
that the peak of the bond-angle distribution is  
sharp at pure limits and broad at or close to the maximum mixing case
($x=0.5$). This is an indication that there are large bond-angle distortion
where the strain is largest. To understand in more detail how the
angular strain affects the local structure, we plotted the 8 partial 
bond-angle distribution functions $g_{\alpha \beta \gamma}(\theta)$
at $x=0.25, 0.5,$ and 0.75, respectively 
as shown in Fig. 6. Here $g_{\alpha \beta \gamma}(\theta)$ 
gives the angular distribution for the angle $\theta$ between two
bonds $\beta \alpha$ and $\beta \gamma$. From Fig. 6, we found that 
all the partial bond-angle distribution functions
$g_{\alpha \beta \gamma}(\theta)$  have their peaks centered  
around the tetrahedral angle ($\theta=109.47^{\circ}$). But both the magnitude
and the position of the peaks change 
as the composition changes and hence so do their 
contributions to the total bond-angle distribution 
function $g(\theta)$. For example,
$g_{\rm SiSiSi}(\theta)$ contributes the most to the total bond-angle
distribution function $g(\theta)$ in the case of $x=0.25$. The magnitude
of its peak shows a monotonical decrease as $x$ increases towards the Ge-rich region.
Simultaneously, the position of its peak shifts from $\theta=109.7^{\circ}$
at $x=0.25$ to $\theta=110.1^{\circ}$ at $x=0.75$. On the other hand, the magnitude of the peak of 
$g_{\rm GeGeGe}(\theta)$ decreases monotonically as $x$ decreases towards the
Si-rich region while the position of the peak shifts from $\theta=109.3^{\circ}$ at $x=0.75$ to 
$\theta=108.8^{\circ}$ at $x=0.25$. In the vicinity of $x=0.5$, however, all of the partial
bond-angle distribution functions $g_{\alpha \beta \gamma}(\theta)$
contribute to the total bond-angle distribution function $g(\theta)$ with
comparable weights.
 
An interesting feature is that the average bond-angles
always keep in the order of
$\bar{\Theta}_{\rm SiSiSi} >\bar{\Theta}_{\rm SiGeSi}> ...
> \bar{\Theta}_{\rm GeSiGe} >\bar{\Theta}_{\rm GeGeGe}$, independent of
the composition (see Fig. 7). 
In particular, it is found that
the deviations of $\bar{\Theta}_{\rm SiSiSi}$ (solid circles)
and $\bar{\Theta}_{\rm SiGeSi}$ (solid squares) from the tetrahedral angle are
always positive while those of $\bar{\Theta}_{\rm GeGeGe}$ (open circles) and
$\bar{\Theta}_{\rm GeSiGe}$ (open squares) always negative. 
This can be understood as follows. 
The average bond-angle $\bar{\Theta}_{\rm SiSiSi} $ ($\bar{\Theta}_{\rm SiGeSi}$)
between two Si-Si bonds (two Ge-Si bonds) depends on how 
the lattice constant of the Si${}_{1-x}$Ge${}_{x}$ alloys at
a certain composition $x$ compares with that of bulk Si. Since the
lattice constant of the alloy is always greater than that of the
bulk Si and it increases with increasing $x$, the average
bond-angle $\bar{\Theta}_{\rm SiSiSi} $ ($\bar{\Theta}_{\rm SiGeSi}$) 
will therefore always be greater than the overall average
bond-angle (almost identical to the tetrahedral angle of 109.47${}^{\circ}$)
and increases with increasing $x$ (see, Fig. 7). Hence, a positive
angular deviation of $\bar{\Theta}_{\rm SiSiSi}$ ($\bar{\Theta}_{\rm SiGeSi}$) 
from the overall average bond-angle results and this angular deviation 
increases with increasing $x$ (towards the Ge-rich region). By the same 
token, since the lattice constant of the alloy is always less than that of
bulk Ge, the average bond-angle $\bar{\Theta}_{\rm GeGeGe}$
($\bar{\Theta}_{\rm GeSiGe}$) will always be less than the overall
average bond-angle and decreases with decreasing $x$.
Thus, a negative angular deviation from the average bond-angle results 
for the average bond-angle $\bar{\Theta}_{\rm GeGeGe}$ 
($\bar{\Theta}_{\rm GeSiGe}$) and this negative deviation
decreases with decreasing $x$ (towards the Si-rich region).
This scenario also indicates that the largest positive angular
deviation for $\bar{\Theta}_{\rm SiSiSi} $ and $\bar{\Theta}_{\rm
SiGeSi}$ (of the order of 1${}^{\circ}$) occurs in the Ge-rich region (large $x$) 
while the largest negative angular deviation for $\bar{\Theta}_{\rm GeGeGe}$
and $\bar{\Theta}_{\rm GeSiGe}$ ($\sim$ 1${}^{\circ}$) occurs in the
Si-rich region (small $x$). But, because the partial bond-angle distribution
function $g_{\rm SiSiSi}(\theta)$ ($g_{\rm SiGeSi} (\theta)$) is
insignificant in the Ge-rich region and $g_{\rm GeGeGe}(\theta)$
($g_{\rm GeSiGe}(\theta)$) is insignificant in the Si-rich
region, the large positive angular deviation of $\bar{\Theta}_{\rm SiSiSi} $ 
($\bar{\Theta}_{\rm SiGeSi}$) and the large negative angular deviation of
$\bar{\Theta}_{\rm GeGeGe}$ ($\bar{\Theta}_{\rm GeSiGe}$) will
not manifest themselves in any significant way in the release
of strain. However, in the neighborhood of $x=0.5$, both the
positive angular deviation of $\bar{\Theta}_{\rm SiSiSi} $ 
($\bar{\Theta}_{\rm SiGeSi}$) and the negative angular deviation
of $\bar{\Theta}_{\rm GeGeGe}$ ($\bar{\Theta}_{\rm GeSiGe}$)
are still substantial ($\sim 0.5^{\circ}$)
while their respective partial bond-angle distribution
functions all make significant contributions to the total
bond-angle distribution.

A closer examination of Figs. 6 and 7
reveals that in the vicinity of $x=0.5$, the magnitudes of
the negative angular deviations of $\bar{\Theta}_{\rm GeGeGe}$ 
and $\bar{\Theta}_{\rm GeSiGe}$ are greater than or comparable
to the positive angular deviations of $\bar{\Theta}_{\rm SiSiSi} $
and $\bar{\Theta}_{\rm SiGeSi}$. Furthermore, the magnitudes of the peaks
of the partial bond-angle distribution functions
$g_{\rm GeGeGe}(\theta)$ and $g_{\rm GeSiGe}(\theta)$ are
greater than those of $g_{\rm SiSiSi}(\theta)$ and $g_{\rm SiGeSi}(\theta)$.
The combination of those effects leads to the situation where the
negative angular deviations outweigh the positive angular deviations.
A net negative angular deviation in the bond-angle manifests itself in the reduction 
of the lattice constant. Hence, in the vicinity of $x=0.5$, it is the
bond angle relaxation that leads to the negative deviation
from Vegard's law in the lattice constant.

Based on the local structural analysis, we have established 
how the local structure changes and how these changes are
the result of the accommodation to the strain relaxation. Even though Si-Si, Si-Ge,
and Ge-Ge pairs in the alloys prefer to maintain their respective bond
lengths so that their respective bond lengths are, for the most part, 
insensitive to the change in the composition, the average overall
bond-length nevertheless shows a linear dependence on the composition. 
The strain relaxation can therefore be separated into two parts: the 
radial relaxation and the angular relaxation. 
The former is responsible for the general trend of a linear dependence on
the composition, and the latter is responsible for the negative deviation
in the lattice constant in the vicinity of $x=0.5$.

\section{Local Analysis of Electronic Structure}

Having identified the mechanism for the strain release associated
with the lattice mismatch and established the link between the
local structural changes and the mechanisms for the strain 
release, it would be illuminating if one can gain an understanding
of the interplay among the local electronic structure, local
structural changes, and the strain-relaxation. For this purpose, we
conducted a local analysis of the electronic structure for
Si${}_{1-x}$Ge${}_{x}$ alloys, using the approach developed
in Ref. \cite{alfonso}. The analysis was carried out in the 
framework of the $sp^{3}$ basis set used in the '{\it ab initio}'
molecular dynamics scheme. In Tables I and II, we list the local
electron distributions (the on-site orbital electrons, the bond electrons, 
and the total site electrons) 
for two alloy configurations corresponding to $x \approx 0.1$ and
$x \approx 0.9$, respectively. Specifically, for each of these
configurations, we present the
results for local electron distributions in two regions: 
(a) one in the vicinity of an impurity atom
and (b) in the vicinity of a host atom (which is at least 
beyond the second nearest neighbors of the impurity atom).
In Table I (corresponding to a Si-rich configuration with
$x=0.1$), the local electron distributions in the region
surrounding a Ge impurity at site 1 with 4 nearest 
neighbor Si atoms at sites 5, 70, 167, and 200, respectively
and those in the region, away from the impurity at site 1,
surrounding a host Si atom at site 213 with 4 nearest
neighbor Si atoms at sites 209, 210, 211, and 212 are given. 
From Table I, it is seen that the charge transfer occurs
between the Ge impurity atom and its 4 nearest neighbor
Si atoms in the region surrounding the impurity while there
is no significant charge transfer amongst the Si atom
and its neighbors in the region away from the immediate
neighborhood of the impurity. From the results shown 
in Tables I and II, it can be seen that the electron transfer 
always occurs from the Si atom to the Ge atom as expected because
of the higher electro-negativity of the Ge atom with respect
to the Si atom. For example, each of the 4 nearest neighbor
Si atoms transfers about 0.017e to the impurity Ge atom
at the center, leading to a gain of about 0.066e for the
Ge impurity atom. The transfer of the electrons comes mostly
from the Si $p$-orbitals to Ge $p$-orbitals, with the 
remainder contributing to the enhancement of the bond charge
for the formation of Ge-Si bond, as can be seen from the
results shown both in Table I ($x$=0.1, Si-rich configuration) 
and Table II ($x$= 0.9, Ge-rich configuration). For the Ge-rich
configuration, the charge transfers again are mainly concentrated in the 
immediate vicinity of the impurity (Si). For example, 
it can be seen from Table II that 0.065e are transfered from
the impurity Si atom (at site 29) to its 4 nearest
neighbor Ge atoms (at sites 25, 26, 27, and 28, respectively),
each gaining on the average about 0.016e. In the region away from
the immediate neighborhood of the impurity Si atom, there
is hardly any charge transfer between the host Ge atom at
site 45 (center) and its 4 nearest neighbor host Ge atoms
(at sites 41, 42, 43, and 44, respectively). From Table I,
one obtains the average bond charge for a pair of Si-Si
bond to be about 0.62e. From Table II, one obtains the
average bond charge for a pair of Ge-Ge bond to be
about 0.58e. From Table I and II, one obtains the average
bond charge for a Si-Ge bond to be about 0.60e. Thus
the sequence of average bond charges follows the order $\bar{q}_{\rm bond}
({\rm Ge-Ge}) < \bar{q}_{\rm bond} ({\rm Si-Ge}) <
\bar{q}_{\rm bond}({\rm Si-Si})$, providing the electronic
basis for the observation of average bond-lengths in the 
order of $b_{\rm GeGe} > b_{\rm SiGe} > b_{\rm SiSi}$
(see section III). Our local analysis also finds that the average
bond charges for the Ge-Ge, Ge-Si, and Si-Si pairs show weak
dependence on the composition. However, the overall average
bond charge shows a general trend of linear dependence on
the composition. These results provide the electronic
basis for the relationships established for various average
bond-lengths in the structural analysis given in section III.

\section{Conclusion}
In conclusion, our {\it ab initio} molecular dynamics study 
of Si$_{1-x}$Ge$_{x}$ alloys provide a comprehensive
understanding of the interplay between the strain relaxation
associated with the lattice mismatch and the changes in the local
structural and electronic properties. We find that Si and Ge
atoms do not have a strong preference to form either the Si-Si or 
the Ge-Ge pair but are fully miscible in the Si$_{1-x}$Ge$_{x}$ alloys 
because of the similar chemical properties.  In the
relaxation process, most of the Si-Si and Ge-Ge pairs try to maintain their
equilibrium distances corresponding to their pure limits, leading to a weak
dependence of the average bond length for Si-Si, Si-Ge, and Ge-Ge pairs on the
composition.  However, the overall average bond-length does show a linear
dependence on the composition (see Fig. 3), indicating that the
radial-relaxation is mainly responsible for the general trend of a linear dependence
of the lattice constant on the composition. On the other hand, the
bond-angle-relaxation in the vicinity of $x=0.5$ has been shown to be responsible for
the negative deviation of the lattice constant from the Vegard's law.\\

\noindent
Acknowledgements\\

This work was supported by the NSF Grants (DMR-9802274 and
DMR-0081006, respectively). We thank Drs. A. Demkov and O. Sankey for the use of 
Fireball96 code. Dr. Yu would like to thank the Institute of Physics, Academia Sinica, Taiwan for 
supporting her stay in Taiwan. We acknowledge the usage of
computing facilities at Dahlem Supercomputing Centre, University of Louisville.

\begin{table}
\caption{The local electron distributions (the on-site orbital 
(s, p${}_{x}$, p${}_{y}$, p${}_{z}$)
electrons, the bond electrons associated with the atom at a given site
with its nearest neighbors, and the total electrons associated with the atom at
a given site) of the alloy Si${}_{1-x}$Ge${}_{x}$ with $x \approx $0.1 
(194 Si atoms and 22 Ge atoms in the supercell). The first column
gives the atomic site label which indicates the position of the atom 
in the supercell (see details in Section IV). Note that the result is obtained
in the framework of $sp^{3}$ basis set.}
\vspace{0.5cm}
\begin{tabular}{|c|c|c|c|c|c|c|}
site label & s & p${}_{x}$ & p${}_{y}$& p${}_{z}$ & bond
electrons & total electrons \\
\hline
 1   (Ge)   &1.36413 &   0.50554 &  0.50491 &   0.50517 & 1.18629 &  4.06604\\
 5   (Si)   &1.23439 &   0.50791 &  0.50881 &   0.50834 & 1.22400 & 3.98344\\
 70  (Si)   &1.23474 &   0.50842 &  0.50851 &   0.50847 & 1.22387 & 3.98401\\
 167 (Si)   &1.23282 &   0.50800 &  0.50844 &   0.50834 & 1.22485 & 3.98245\\
 200 (Si)   &1.23460 &   0.50819 &  0.50884 &   0.50902 & 1.22405 & 3.98470\\
 209 (Si)   &1.23520 &   0.51266 &  0.51200 &   0.51179 & 1.23056 & 4.00220\\
 210 (Si)   &1.23458 &   0.51178 &  0.51219 &   0.51240 & 1.23107 & 4.00202\\
 211 (Si)   &1.23539 &   0.51216 &  0.51181 &   0.51152 & 1.23084 & 4.00171\\
 212 (Si)   &1.23517 &   0.51218 &  0.51267 &   0.51212 & 1.23047 & 4.00261\\
 213 (Si)   &1.23523 &   0.51142 &  0.51086 &   0.51186 & 1.22959 & 3.99896 \\
\end{tabular}
\end{table}  

\begin{table}
\caption{The local electron distribution (the on-site orbital (s,
p${}_{x}$, p${}_{y}$, p${}_{z}$)electrons, the bond electrons 
associated with the atom at a 
given site with its nearest neighbors, the total electrons associated with the atom at
a given site) of the alloy Si${}_{1-x}$Ge${}_{x}$ with $x \approx $0.9 
(22 Si atoms and 194 Ge atoms in the supercell). The first column gives the site label
which indicates the position of atom in the supercell (see details in Section IV).
Note that the result is obtained
in the framework of $sp^{3}$ basis set.}
\vspace{0.5cm}
\begin{tabular}{|c|c|c|c|c|c|c|}
site label & s & p${}_{x}$ & p${}_{y}$& p${}_{z}$ & bond
electrons& total electrons\\
\hline
25 (Ge)& 1.37737 & 0.49334 & 0.49340 & 0.49316& 1.15640  & 4.01367\\
26 (Ge)& 1.37615 & 0.49374 & 0.49376 & 0.49358& 1.15781  & 4.01504\\
27 (Ge)& 1.37787 & 0.49430 & 0.49424 & 0.49440& 1.15614  & 4.01695\\
28 (Ge)& 1.37739 & 0.49370 & 0.49358 & 0.49447& 1.15631  & 4.01544\\
29 (Si)& 1.24569 & 0.49890 & 0.49896 & 0.49910& 1.19274  & 3.93539\\        
41 (Ge)& 1.37833 & 0.49107 & 0.49134 & 0.49121& 1.14972  & 4.00167\\
42 (Ge)& 1.37836 & 0.49046 & 0.49051 & 0.49103& 1.15002  & 4.00037\\
43 (Ge)& 1.37892 & 0.49140 & 0.49139 & 0.49130& 1.14951  & 4.00252\\
44 (Ge)& 1.37801 & 0.49069 & 0.49059 & 0.49041& 1.15051  & 4.00021\\ 
45 (Ge)& 1.37814 & 0.49045 & 0.49013 & 0.49083& 1.14994  & 3.99949\\
\end{tabular}
\end{table}

\newpage
\noindent{\bf FIGURE CAPTIONS}
\vskip 0.1 in

\noindent{Fig. 1: The calculated coordination numbers as a function of
the composition $x$
where the circles denote $N_{\rm SiSi}$, the up-triangles 
$N_{\rm SiGe}$, the down-triangles $N_{\rm GeSi}$, and
the squares $N_{\rm GeGe}$. The solid symbols correspond to
the {\it random} type and the open ones correspond to the
{\it regular} type (see the text for definitions).  
The inset shows the coordination numbers obtained from XAFS
results
as a function of composition {\protect \cite{aubry}}.
The error bars in Fig. 6 of Ref. {\protect \cite{aubry}} are not
shown here for the sake of presenting a clean comparison the
trend of $N_{\alpha \beta}$ between the theoretical results
and the experimental measurements. The lines indicate
the coordination numbers predicted from x-ray diffraction
assuming random 
site occupancy.}
 \vskip 0.1 in
\noindent{Fig. 2: Optimized lattice constant as a
function of the composition $x$ (solid circles). The solid line
is the
Vegard model prediction. The inset is the experimental
measurements where 
solid circles are from Ref.{\protect \cite{expt}} and open
circles
are from Ref.{\protect \cite{aubry}}. Note that the error bars in
Fig. 7
of Ref. {\protect \cite{aubry}} are not shown in the inset for
the sake of
presenting a clean comparison of the trend between the
theoretical and
the experimental results.}
\vskip 0.1 in
\noindent{Fig. 3: Calculated average bond-lengths as a function of the
composition $x$, 
where the circles denote $b_{\rm SiSi}$, the up-triangles
$b_{\rm SiGe}$, and the squares $b_{\rm GeGe}$. The inset is the
experimental result for the first-shell bond-lengths at different
compositions (see Fig. 8 in Ref.{\protect \cite{aubry}}). Note
that
the error bars in Fig. 8 of Ref. {\protect \cite{aubry}} are not
shown in the inset because our purpose is to compare the trend
of the bond length vs composition between the theoretical
calculation
and the experimental observation. The stars 
represent the overall average bond-lengths calculated by taking
into account all three types of bonds without distinguishing any 
particular bonding pair. A solid line is drawn through these
points to provide a guidance to the eye.}
\vskip 0.1 in
\noindent{Fig. 4: The total $g(r)$ (solid curves) and partial $g_{\alpha
\beta}(r)$
(dashed curves) radial-pair distribution function at compositions
of
$x$=0.0, 0.10, 0.25, 0.40, 0.5, 0.60, 0.75, 0.90, and 1.0.
Note that the left dashed curves correspond to
$g_{\rm SiSi}(r)$, the middle to $g_{\rm SiGe}(r)$, and the right
to $g_{\rm GeGe}(r)$, respectively. } 
\vskip 0.1 in
\noindent{Fig. 5: The total bond-angle distribution function $g(\theta)$
at $x$=0.0, 0.10, 0.25, 0.40, 0.5, 0.60, 0.75, 0.90,and 1.0.} 

\noindent{Fig. 6: The partial bond-angle distribution
function 
$g_{\alpha \beta \gamma}(\theta)$ at $x=0.25, 0.5$, and $0.75$.
The left panel shows the results corresponding to $g_{\rm
SiSiSi}(\theta)$ (solid line),
$g_{\rm SiGeSi}(\theta)$ (dotted line), $ g_{\rm GeSiSi}(\theta)$
(dashed line), and $g_{\rm SiSiGe}(\theta)$ (long-dash line). The
right panel shows the results corresponding to $g_{\rm
GeGeGe}(\theta)$ (solid line), $g_{\rm GeSiGe}(\theta)$ (dotted
line),
$g_{\rm SiGeGe}(\theta)$ (dashed line), and $g_{\rm
GeGeSi}(\theta)$ (long-dash line).} 
\vskip 0.1 in
\noindent{Fig. 7: The average bond-angle $\bar{\Theta}_{\alpha \beta
\gamma}$
as a function of the composition $x$. Note that the stars denote
the 
overall average bond-angle, the solid (open) circles
$\bar{\Theta}_{\rm SiSiSi}
(\bar{\Theta}_{\rm GeGeGe})$, the solid (open) squares
$\bar{\Theta}_{\rm SiGeSi} \; (\bar{\Theta}_{\rm GeSiGe})$,
the solid (open) up-triangles $\bar{\Theta}_{\rm GeSiSi} \;
(\bar{\Theta}_{\rm SiGeGe})$, and the solid (open) down-triangles
$\bar{\Theta}_{\rm SiSiGe} \; (\bar{\Theta}_{\rm GeGeSi})$,
respectively. The solid line denotes the tetrahedral angle of
109.47$^{\circ}$.}
 
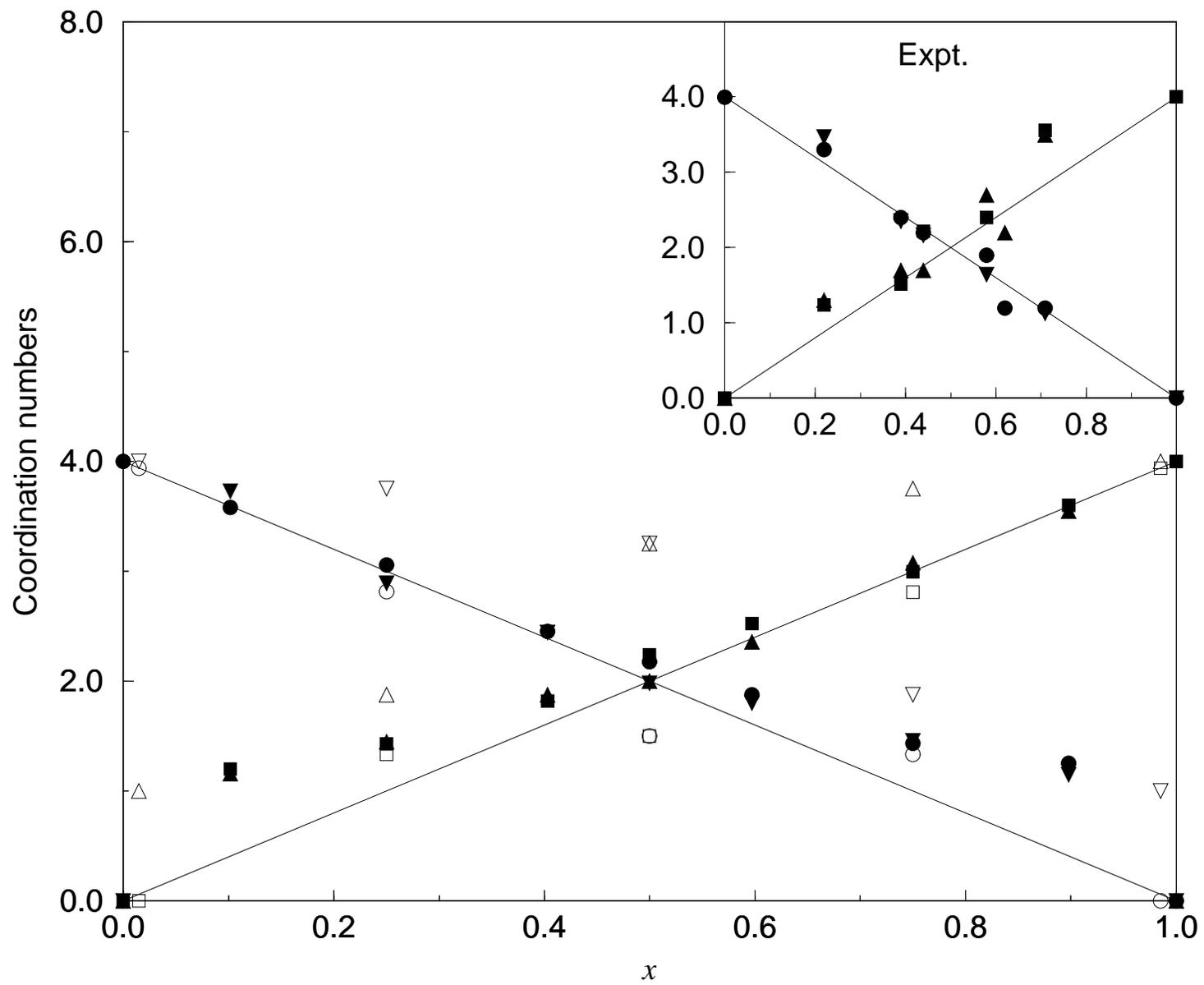
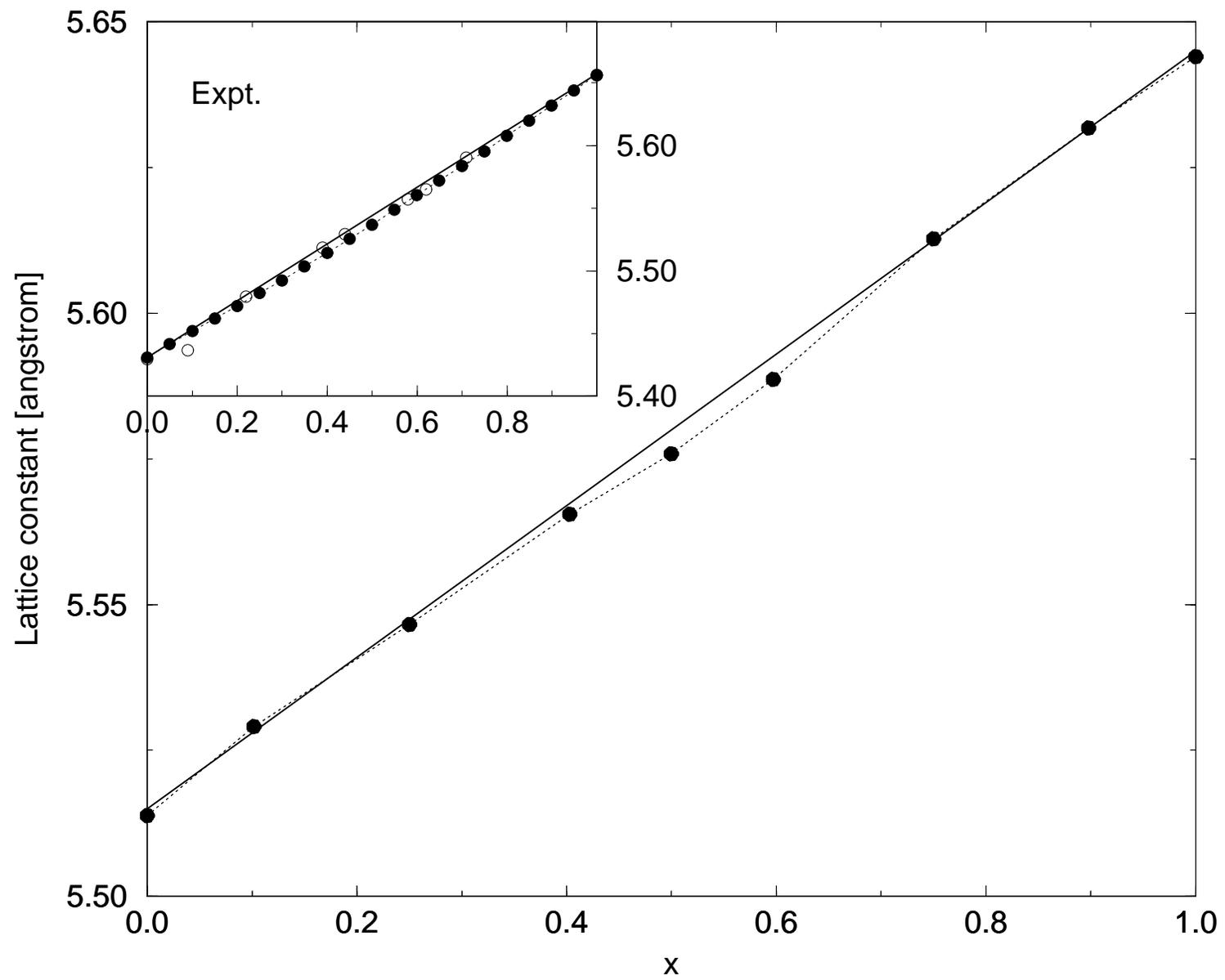
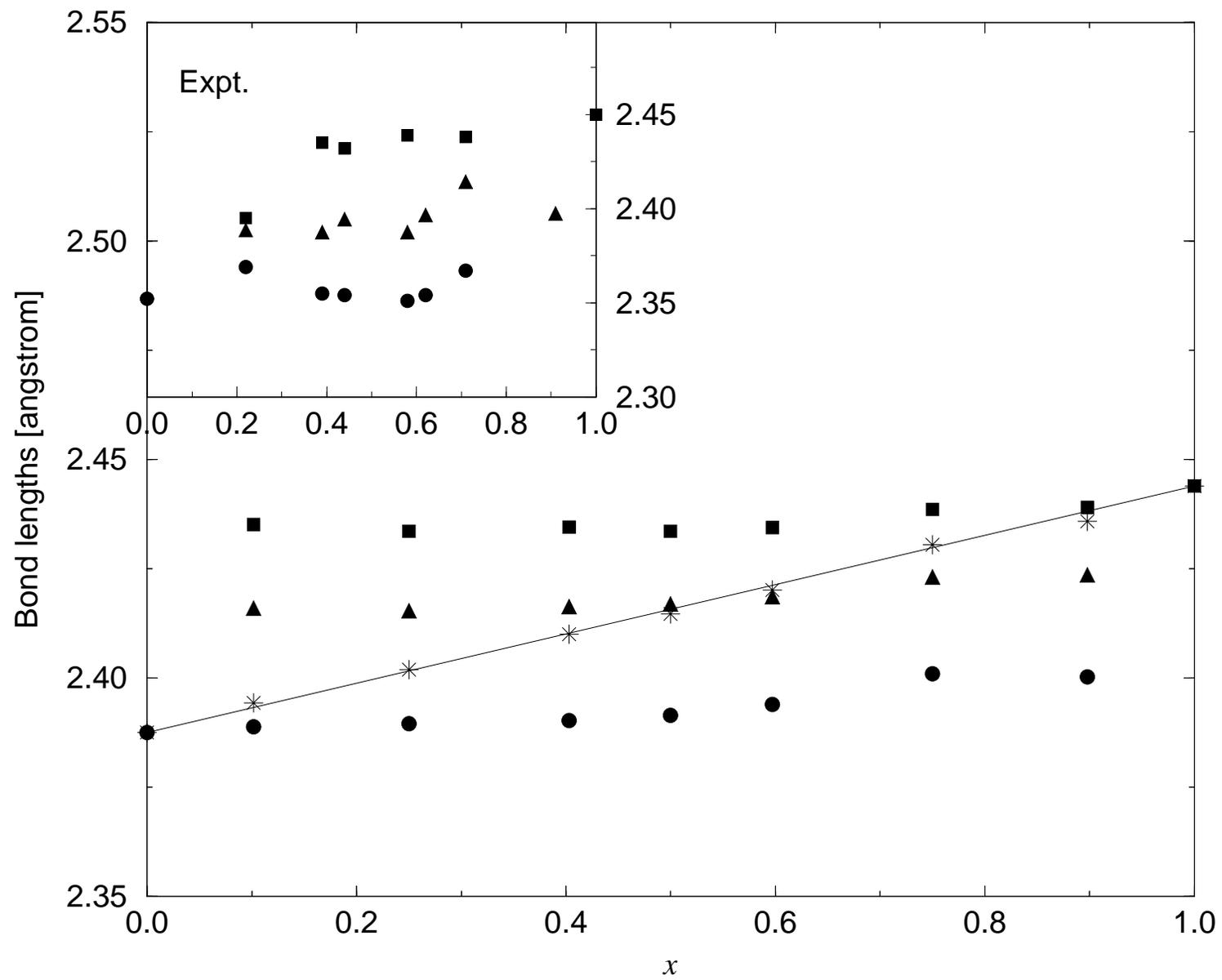
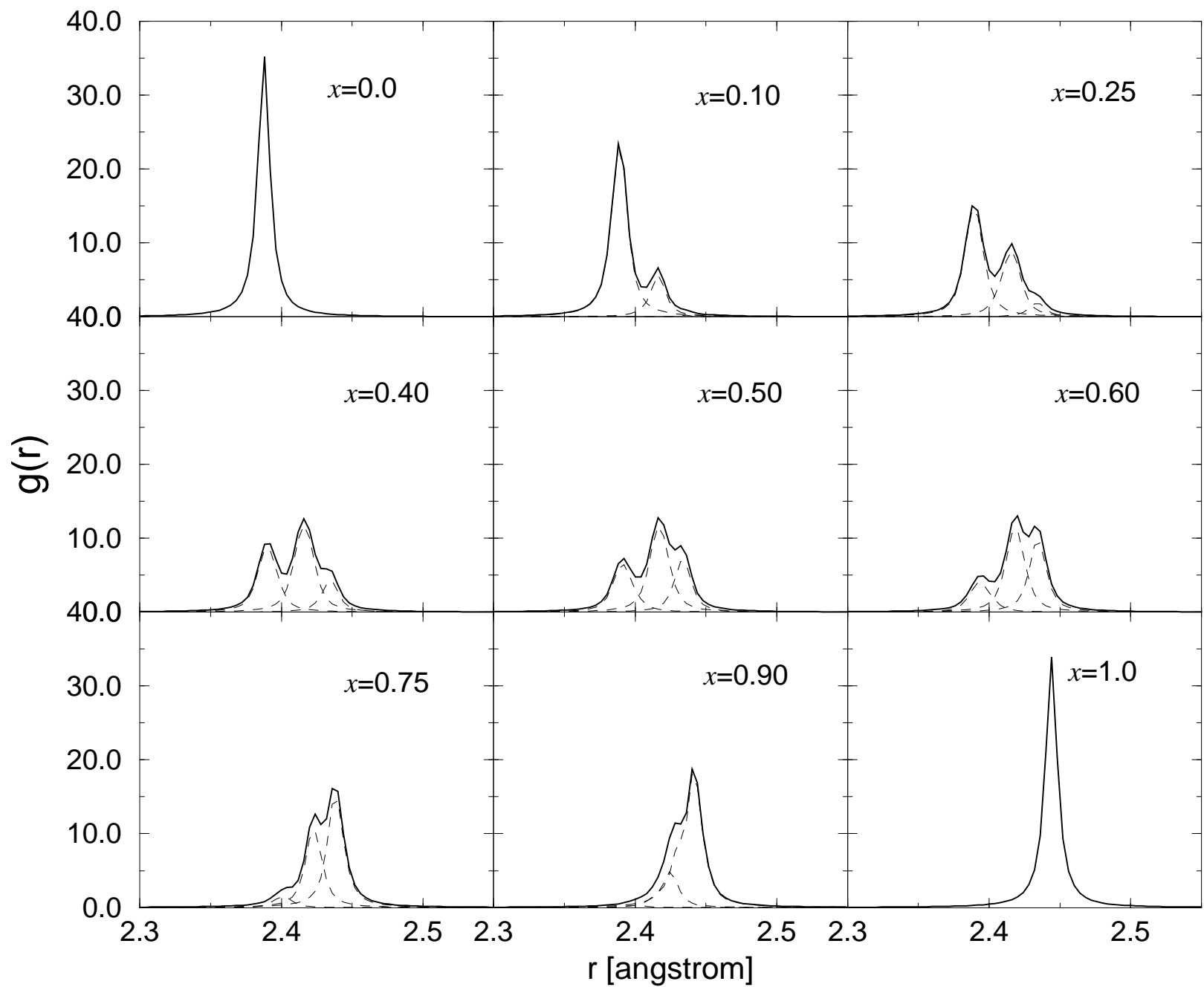
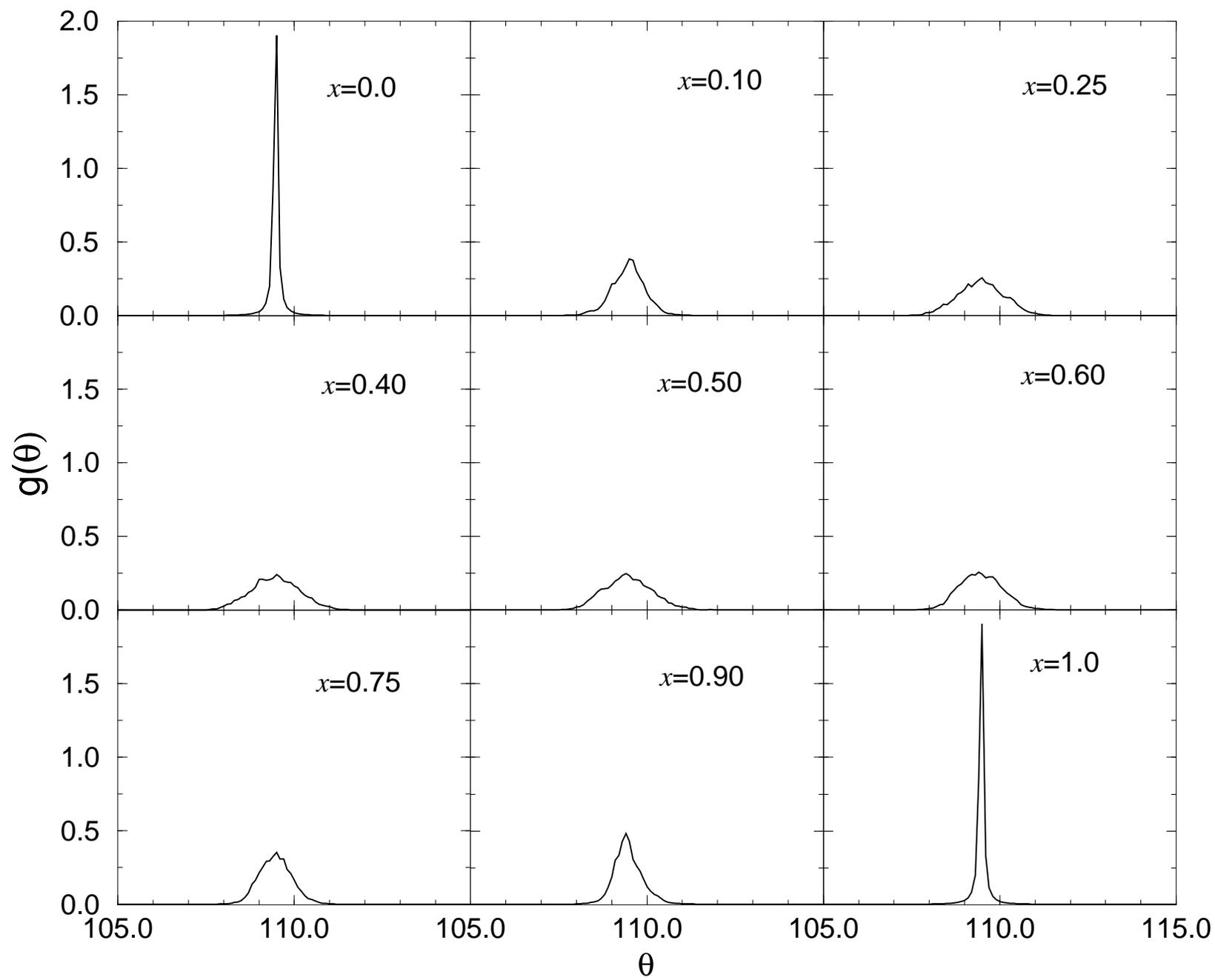
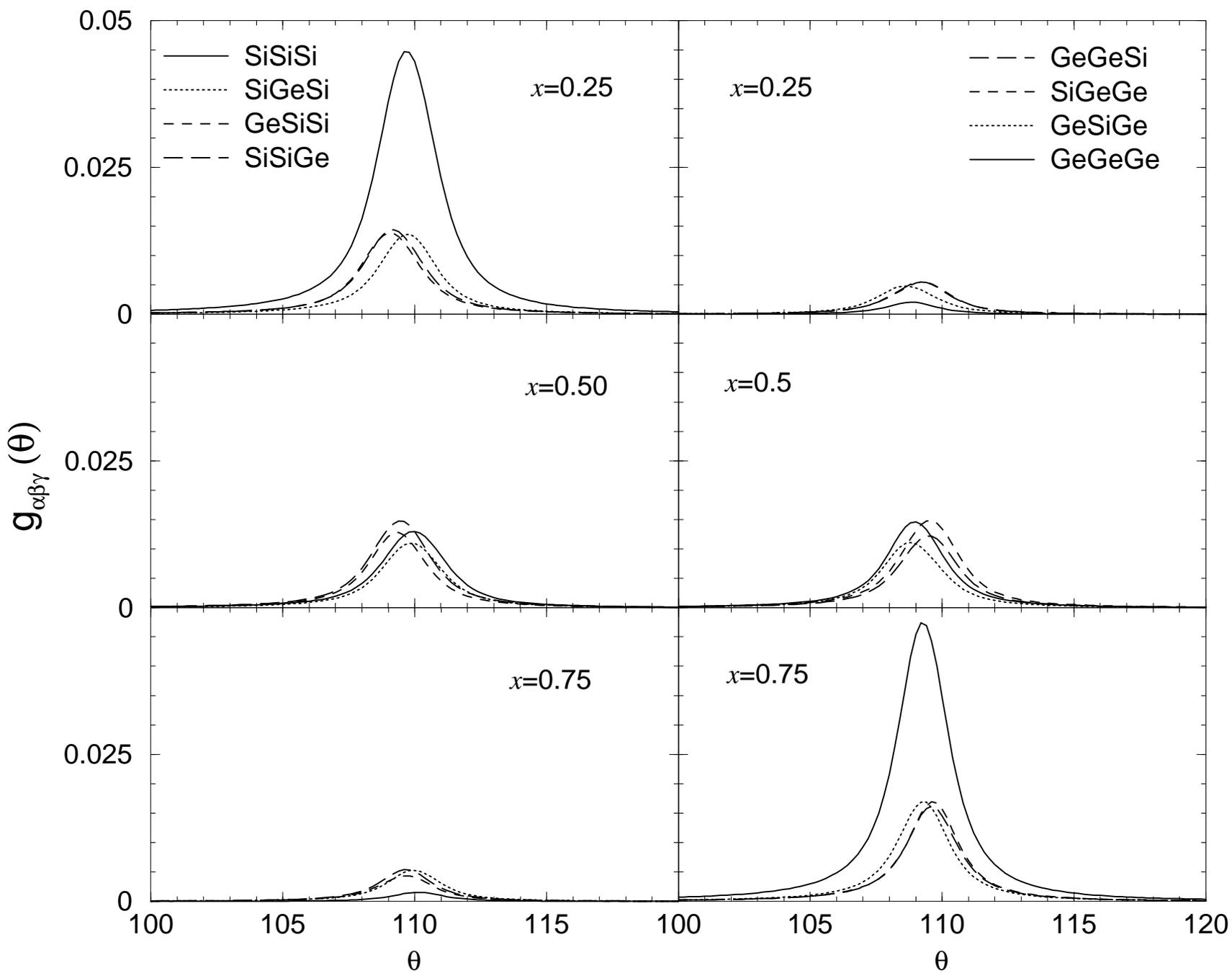
\end{document}